# Incorporating a 'ladder of trust' into dynamic Allocation of Function in Human-Autonomous Agent Collectives

Chris Baber[1*], Patrick Waterson[2], Sanja Milivojevic[3], Sally Maynard[2], Edmund Hunt[3], Sagir Yusuf[1],
1University of Birmingham, 2Loughborough University, 3University of Bristol
e-mails: c.baber@bham.ac.uk; P.Waterson@lboro.ac.uk; sanja.milivojevic@bristol.ac.uk;
S.E.Maynard@lboro.ac.uk; Edmund.Hunt@bristol.ac.uk; s.m.yusuf@bham.ac.uk

**Abstract**

A major, ongoing social transition is the inclusion of autonomous agents into human organizations. For example, in defence and security applications, robots may be used alongside human operatives to reduce risk or add capability. But a key barrier to the transition to successful human-autonomous agent collectives is the need for sufficient trust between team members. A critical enabling factor for this trust will be a suitably designed dynamic allocation of function (AoF). We consider AoF in terms of a 'ladder of trust' (from low to high) with individual team members adjusting trust in their teammates based on variation in 'score' over time. The score is derived by the ability of team member to perceive and understand its situation based on the gathered information and act to acheive team or self goals. Combining these trust scores gives a system-level perspective on how AoF might be adjusted during a mission. That is, the most suitable teammate for a function might have a low trust rating from its fellow teammates, so it might be preferable to choose the next most suitable teammate for the function at that point in time. Of course, this is only in the situation where the next most suitable teammate is also likely to perform within the set framework of moral, ethical, and legal constraints. The trade-offs between trust in the individual agent's capability and predictability need to be considered within the broader context of the agent's integrity and accountability. From this perspective, the Allocation Space is defined by more than ability of each agent to perform a function. The models that we are developing also require cooperation (and communication) between agents. This can allow the proposed AoF to be negotiated between agents and leads to the proposal that AoF could, in effect, represent a 'contract' between the agent performing the function and the agents that would be affected by this performance. We argue that this new approach to trust-sensitive AoF could be an important enabler for organizations seeking to embrace the opportunities arising from integrating autonomous agents into their teams.

*Introduction*

Autonomous agents are increasingly integrated into work organizations. Yet, human trust in the autonomous agents is a key barrier to their acceptance and effective use. The division of tasks between humans and machines in complex systems has traditionally been referred to as Allocation of Function (AoF) (Older et al. 1997). The initial contribution to AoF was made by Fitts et al. (1951) who produced a list of tasks for both the human and the machine on which the performance of one exceeds that of the other. This and later versions are generically referred to as 'Fitts' Lists' ('Humans-are-better-at / Machines-are-better-at, HABA-MABA). Contemporary approaches to AoF emphasize the wider sociotechnical context and include organizational, teamworking, and job design criteria (Waterson et al. 2002).
Jennings et al. (2014) discuss how human-agent collectives (HACs) have become ubiquitous in everyday life. HACs are a new class of socio-technical systems in which humans and smart software (agents) engage in flexible relationships in order to achieve both their individual and collective goals. In HACs, sometimes the humans take the lead, sometimes the computer (or robot) does, and this relationship can vary dynamically. More recently, Roth et al. (2019) reviewed AoF methods in the light of human-autonomy teaming and concluded that AoF is fundamentally about exploring a trade-space of alternative AoF options. An effective AoF method should identify these trade-offs. A key element of

# Incorporating a 'ladder of trust' into dynamic Allocation of Function in Human-Autonomous Agent Collectives

such trade-offs is the importance of issues related to job design, human-human AoF (Stemmers and Hallam, 1985) and the wider importance of trust. Another requirement is that AoF needs to move beyond the 'who does what' approach to allocation, focussing not just on actions alone, but moving towards a more nuanced viewpoint where ethics and cooperations are key elements (Dekker and Woods, 2002). Asaro and Wallach (2017) note, we need to consider what capabilities are required for autonomous systems to recognise when they are in ethically significant situations, and how we can factor human ethical concerns into selecting safe, appropriate and moral course of action in HACs. Furthermore, as Chiou and Lee (2021) have recently pointed out, contemporary approaches to trust have often overlooked questions of cooperation between agents. As agents become capable of working autonomously and in pursuit of their own goals, then the question of trust becomes more important.

In this paper, we explore a 'trust space' through a simple scenario and revisit some of the questions relating to AoF in the era of HACs. Following Lewis and Marsh (2022), we consider 'trust' in terms of predictability, capability, and integrity. Capability is describes whether an agent (or human) is able to perform a given function at a given point in time (in its own or its team-mates opinion). This could be defined from its capability (i.e., whether it can complete the function with or without assistance) and its availability (i.e., whether it is in the appropriate location and whether it is currently occupied). Predictability involves the probability of success of completing the function. We see integrity as incorporating the tension between pursuing individual versus team goals, as well as maintaining the performance/actions within the set of moral, legal, and ethical constraints set by the mission control or other leading agent(s), which can impact on the resilience of the team. In this way, 'trust' can be considered in terms of a 'model' of teammates held by members of a team in which each component can be adjusted in the face of new information. Thus, 'trust' is not a static concept but can adapt to ongoing team performance.

*Methods*

To explore AoF in a HAC (where agents can have different capabilities, across varying situations and goals defined by environment, location, acquired information and interpretations, and roles), we use a simple maze exploration scenario. The justification for this scenario is that it relates to some of the Security patrol and Military search activities raised by our stakeholders but is sufficiently abstract and simple to allow computer modelling. It exemplifies our operationalization of the trust concept per the Lewis and Marsh (2022) framework.

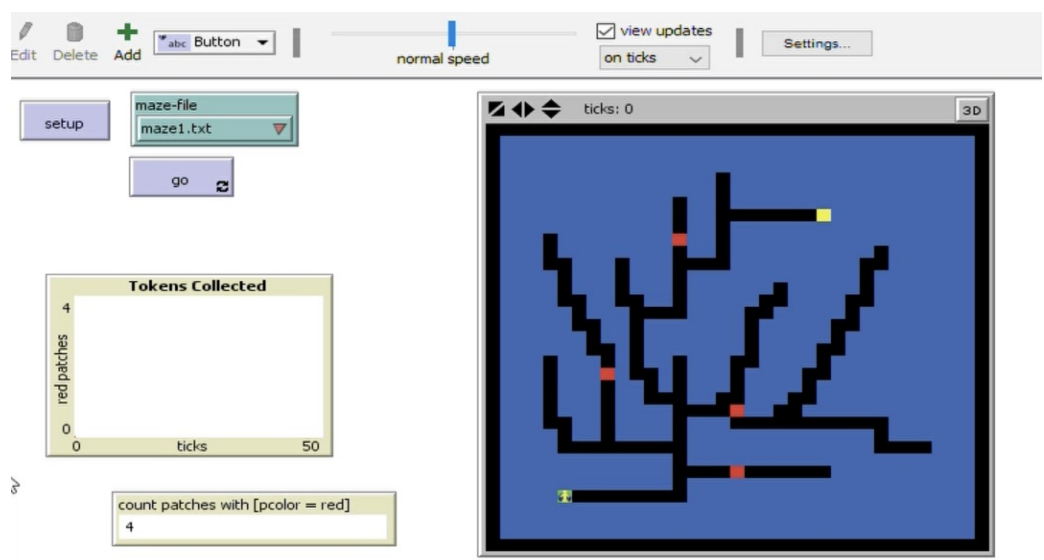

Figure 1: NetLogo model of maze task

Maze Task

In a NetLogo Agent-based simulation (Figure 1), four agents (positioned in the bottom left of the maze) explore a maze (following left- or right-hand-on-wall path, i.e., walking through the maze and keeping the left or right hand on the wall at all times). The maze contains 'tokens' (red patches) which specialised agents will collect. However, for other agents the red patches are 'traps' and can only be deactivated when another agent changes the colour of the 'traps'. Agents move through maze and collect tokens, or release agents stuck in traps, or go through as many gates as possible, or follow a leader. Therefore, the agents operation involves detect-reason-act, where detect inolves environment perception, reason involves decisions to ensure goal achievement, and action involves triggered activities (see table 3). In one version of the scenario, there is a Team Goal which is to find the shortest route whereby all team members escape the maze. The easiest way to achieve this is for there to be a leader agent and the other agents follow this. However, agents might have Individual Goals which compete with this Team goal, e.g., entering gates or collecting tokens could be a selfish goal; releasing a Teammate could be an altruistic goal.

Defining a Mission

We use Cognitive Work Analysis (Jenkins et al., 2008; Vicente, 1999) to define the mission. In Cognitive Work Analysis, a system is described through five views. Figure 2 shows the Abstraction Hierarchy, from the Work Domain Analysis view, for the maze task. Here, the overall mission (i.e., functional purpose) is to solve the maze. In an Abstraction Hierarchy, these constraints are defined as *values and priority measures* that the system is intended to achieve. Completing the overall mission involves satisfying several values and priorities measures, e.g., minimize time might conflict with maximize tokens (because collecting tokens might involve at least one agent detouring from the shortest route). Individual agents can use objects in the environment to perform tasks (i.e., *object-related functions*) and task performance contributes to achieving goals (i.e., *purpose-related functions*).

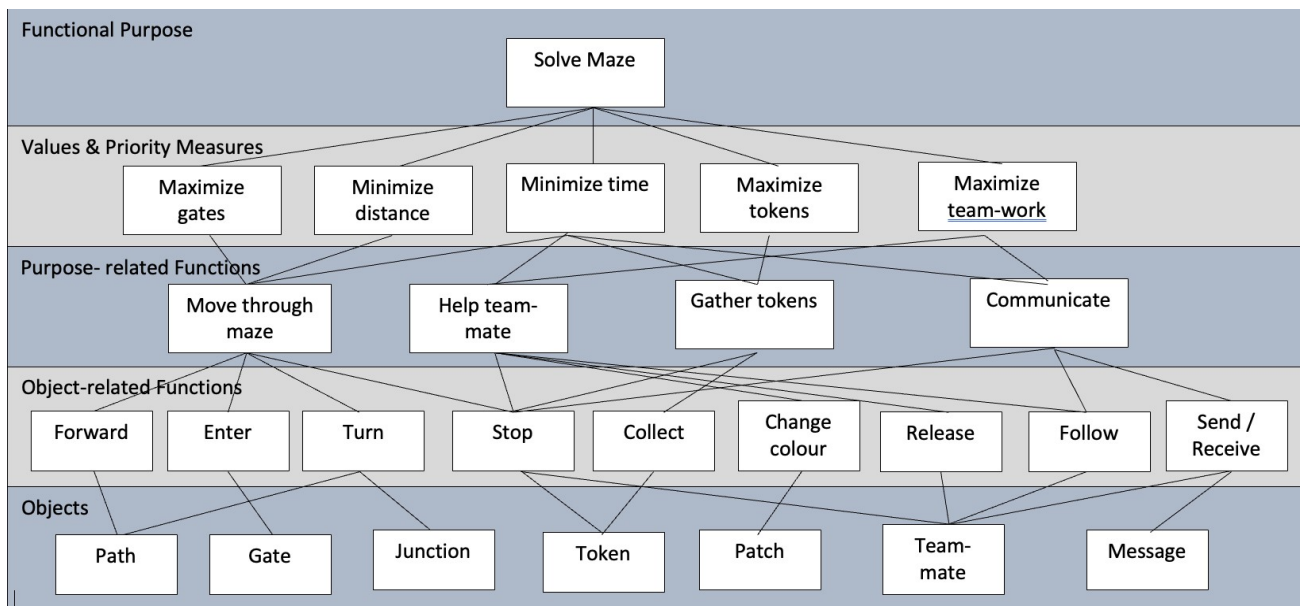

Figure 2: Abstraction Hierarchy of the Maze Task

Defining 'integrity' for team members

Our definition of 'integrity' assumes that the *functional purpose* of the team is well defined and the value and priority measures define the constraints under which the team is operating. These constraints will not only be defined within the boundaries of the Mission but also in terms of safety, legal, ethical, and societal limits to which the team operates. 'Integrity' mandates that a team mate's activity does not violate any of the values and priority measures. That is, an agent might pursue activity that maximizes a value and priority measure that is unique to its role or might pursue activity that maximizes team goals.

# Incorporating a 'ladder of trust' into dynamic Allocation of Function in Human-Autonomous Agent Collectives

Defining Capability

In terms of AoF, we could specify each agent to have a role that involves the capability to perform actions (i.e., object-related functions). From this, AoF could be defined by the assignment of object-related functions before the mission the Social Cooperation and Organization Analysis (SOCA) from CWA. An example of this is shown in Figure 3. In a previous application of CWA to Allocation of Function, Jenkins et al. (2008) used purpose-related functions as the basis of the allocation. In this section, our focus is on object-related functions. However, we note that it is possible that the allocation could made at either level (or, for that matter, at the value and priority measures level). SOCA can reflect the system designers' intentions behind the AoF.

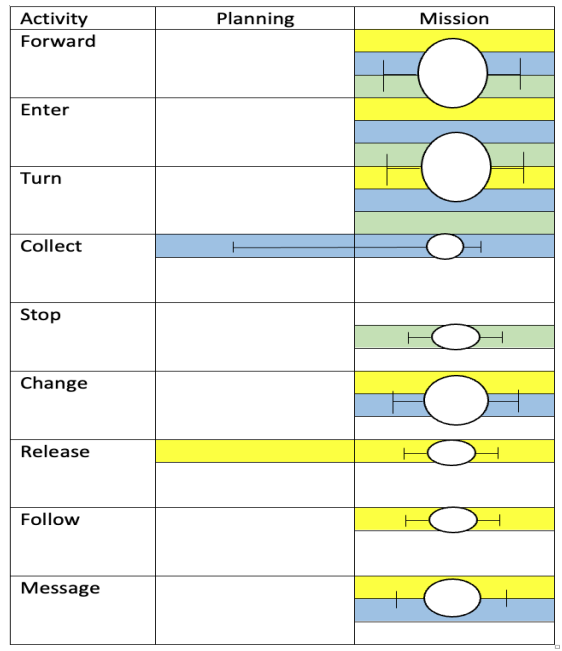

Figure 3: SOCA analysis showing pre-defined roles. Each role is indicated by a different colour, ordered in 3 rows for each activity.

Alternatively, we could give each agent a probability of choosing one action over another, e.g., if the agent sees a token, it could have a x% to move towards that token and collect it and y% to ignore the token. Similar probabilities would relate to choosing to enter a gate or choosing to rescue a trapped team-mate. Table 1 describe an agent's actions using conditional probability table (a table containing all possible outcomes of a situation of an agent). The assigned probabilities can change based on the situation of the agents at any given time and the achievement of both self and team goals. The probability values (for both agents' situation and actions) can be derived from agents' sensor (e.g., location based on GPS), Subject Matter Expert(SME) assignment, or agents training (Yusuf and Baber, 2022).

Table 1: An example of Agent Actions CPT

| Agents Situation Probabilities | | | Agent Actions Probabilitites | |
|---|---|---|---|---|
| Location Situation | Time | Mission Goal | Movement | Activity |
| Cell 1:100% | T1:100% | Maximise Teamwork:80%, Minimise gate:10%, Maximise tokens:10% | Forward:100% | Change color:100% |
| Cell 2:100% | T2:100% | Minimise gate | Backward:100% | Collect:100% |
| Cell 3:100% | T3:100% | Minimise time | Left:100% | Release:100% |
| Cell 4:100% | T4:100% | Maximise tokens | Right:100% | Follow:100% |
| … | … | … | … | … |
| Cn:100% | T n:100% | Minimise/maximise n:100% | Movement n:100% | Activity n:100% |

# Incorporating a 'ladder of trust' into dynamic Allocation of Function in Human-Autonomous Agent Collectives

Defining predictability

While the previous shows how agents can be assigned a purpose-related function, we might also need to consider how well the agent can do this and whether it might require assistance. We consider this through CoActive Design (Johnson et al., 2014). This decomposes system activity into functions and defines the capability of each agent (human or automation) to propose which would be most appropriate. In a sense, this follows the HABA-MABA approach from Fitts' List. Where CoActive Design offers an advantage is that it can also consider the level of confidence each agent might have in performing a function, e.g., in terms of whether the agent requires or can provide assistance for a function. For this paper, the focus is on the predictability of an agent to fulfil a goal, i.e., whether an agent is able to complete a purpose-related function.

Table 2: Colour coding for CoActive Design

| colour | Performer | Supporter |
|---|---|---|
| green | I can perform the function without help | My support can improve efficiency |
| yellow | I can perform the function but my reliability is <100% | My support can improve reliability |
| orange | I can perform some aspects of the function but need support | My support is required |
| red | I cannot perform this function | I cannot provide support for this function |

In our version of CoActive Design (Baber and Vance, 2019), the colours in Table 2 have been scored on an ordinal scale (green = 3; yellow = 2; orange = 1; red = 0) to produce a 'total score' for each team member (table 3). We assign these scores to reflect assumptions about the capability of team members for specific tasks. In this way, the SOCA diagram in Figure 3 can be scaled to reflect differences in agent capability. For example, we could make assumptions about the team members that have higher (or lower) levels of automation or training which can be reflected in their capability. A lower score indicates that the team member might be less flexible in the mission.

Table 3: Definition of Agent Roles

| Purpose-related function | Performers | | | | Supporters | | | |
|---|---|---|---|---|---|---|---|---|
| | Leader | Collector | Gate-user | Neutral | Leader | Collector | Gate | Neutral |
| move through maze | 3 | 3 | 3 | 3 | 0 | 0 | 0 | 0 |
| help team mates | 0 | 1 | 0 | 3 | 0 | 1 | 0 | 3 |
| gather tokens | 0 | 3 | 0 | 0 | 0 | 0 | 0 | 0 |
| communicate | 3 | 2 | 2 | 3 | 2 | 2 | 2 | 2 |
| | 6 | 9 | 5 | 9 | 2 | 3 | 2 | 5 |

Communication between agents

In addition to scoring the individual team members, Baber and Vance (2019) consider how information could be shared between agents, and whether this involves translating from one representation to another. Given that this would be a turn-based game, an agent can choose to move or send a message. The message could be to request help if the agent is in a 'trap' or could be a sighting of a 'token' (if the agent is not collecting the tokens), or a 'follow me' message, or a message to say that the agent has stopped (e.g., to collect a token or because it is trapped). However, as Table 4 illustrates, the environment might yield the same or different interpretations, depending on the agents. It might be the same for all agents (e.g., a black square allows the agent to move forward and a blue square is not passable). In this case, observing another agent on a black square means that their movement can be predicted. In other cases, information, e.g., a red square, has a different meaning for each of the agents. This means that 'common ground' is no longer apparent and there needs to be a 'translation' (i.e., an agent needs to specify its interpretation of that information) in order to enable communication. The act of communicating (and the associated act of translation) might also be a function that an agent chooses to perform or not – particularly if such communication requires effort at the expense of another action.

# Incorporating a 'ladder of trust' into dynamic Allocation of Function in Human-Autonomous Agent Collectives

Table 3: How each Agent might interpret the environment for this scenario

| Activity | Leader | Collector | Gate-user | Neutral |
|---|---|---|---|---|
| Forward | Black and blue squares | Black and blue squares | Black and blue squares | Black and blue squares |
| Enter | Entrance on left or right (depending on capability) | Entrance on left or right (depending on capability) | Entrance on left or right (depending on capability) | Entrance on left or right (depending on capability) |
| Turn | Path to left or right (depending on capability) | Path to left or right (depending on capability) | Path to left or right (depending on capability) | Path to left or right (depending on capability) |
| Collect | - | Red square | Entrance available | - |
| Stop | Red square | Leader stops | Red square or Leader stops | Leader stops |
| Change | - | Square turned to black; Token collected | - | - |
| Release | - | - | Square turned to black; Leader released | - |
| Follow | Look for 'neutral' agent | - | - | Follow leader |
| Message | 'follow me', 'help', 'stop_all' | 'help' (from leader) | 'help' (from leader) | 'follow me' (from leader) |

*Discussion/perspectives*

In this paper, we have used a tripartite definition of 'trust' as capability, predictability, integrity and shown how two Human Factors methods (CWA and CoActive Design) can illustrate this definition. As agents are given different roles or different goals, so the challenges of predictability (in terms of knowing what a teammate is doing) and integrity (in terms of the trade-off between selfish, altruistic, team goals, including moral, ethical and legal frameworks/boundaries) become more apparent. As we consider HACs in a socio-technical reality, it is the system as a whole that acts and reacts, achieves or fails to achieve goals, and succeeds or fails in the set mission. Ultimately, within the rules set out for the team, we want all agents – or more precisely, the system itself - to be accountable for actions (drawing on Actor-Network theory and Haraway's concept of companion species).Thus, for example, if a human agent becomes too reliant on automated actors in the network and, as a consequence, changes behaviour (e.g. becomes less attentive to details, fails to identify risks, prioritises non-altruistic goals, etc.), the overall outcome (a failed mission) is a consequence of a shared failure of the HAC, and accountability is also shared (Kruijff and Janicek 2011). We want to explore how we can develop HACs that will be implicated in one another in caring and complicated ways (Haraway 2016), based on a scaleable ladder of trust that will result in systems exceeding the capacity of humans or robots alone.

In our simulated example, as the Mission progresses, the agents have to make decisions based on their role, the state of their immediate environment, and the goals which are available to them. Running the simulation under different configurations illustrates the combination of decisions, actions and outcomes that are possible and highlights the manner in which AoF can develop dynamically. Using a composite measure of trust (based on the combined measures of capability, predictability, and integrity within the team) reveals that this moves up and down as our concept of a 'ladder of trust' assumes.

One question that this approach raises is why a team member would not seek to pursue team goals. That is, under what circumstances would a teammate behave in a selfish manner. We answer this in three ways. First, autonomous agents might be designed to pursue their own goals in the first instance, where an environment has work that can usefully be done by only one agent. In this case, pursuit of team goals could be viewed as a demand on the agent. Second, the definition of purpose-related functions (goals) might solely be based on an agent's capability, and not take into account the wider constraints. Third, team goals might be affected by constraints which had not been predicted in the initial specification of the functions. Each of these 'selfish' behaviours implies the need to adjust the goals to which team members are working, particularly in terms of team goals. In the absence of team goals, it is plausible to assume that individual agents continue to pursue goals which are relevant to them.

# Incorporating a 'ladder of trust' into dynamic Allocation of Function in Human-Autonomous Agent Collectives

**References**

Asaro, P. and Wallach, W. (2017) The Emergence of Robot Ethics and Machine Ethics. In Wallach, W. and Asaro, P. (eds.) *Machine Ethics and Robot Ethics*. London and New York: Routledge.

Baber, C. and Vance, C. (2019). Design of human-machine teams using a modified CoActive Design Method, In R. Charles & D. Golightly (eds.) *Contemporary Ergonomics 2019*, CIEHF.

Chiou, E.K. & Lee, J.D. (2023). Trusting automation: designing for responsivity and resilience, *Human Factors,* 65, 137-165.

Dekker, S. W. A., & Woods, D. D. (2002). MABA-MABA or abracadabra? Progress on human-automation co-ordination. *Cognition, Technology & Work,* 4, 240–244.

Fitts, P. M., Viteles, M. S., Barr, N. L., Brimhall, D. R., Finch, G., Gardner, E., & Stevens, S. S. (1951). *Human engineering for an effective air-navigation and traffic-control system*, Columbus: Ohio State University Research Foundation.

Haraway, D. (2015) Manifestly Haraway. Minneapolis: University of Minnesota Press.

Jenkins, D.P., Stanton, N.A., Salmon, P. and Walker, G.H. (2008). *Cognitive Work Analysis: coping with complexity*, Avebury: Ashgate

Jenkins, D.P., Stanton, N.A., Salmon, P.M., Walker, G.H. and Young, M.S. (2008). Using cognitive work analysis to explore activity allocation of function within military domains, *Ergonomics, 51,* 798-815.

Jennings, N. *et al.* 2014. Human-Agent Collectives. *Communications of the ACM*, 57, 12, 80-88.

Johnson, M., Bradshaw, J.M., Feltovich, P.J., Jonker, C.M., Van Riemsdijk, M.B. and Sierhuis, M. (2014). Coactive design: Designing support for interdependence in joint activity, *Journal of Human-Robot Interaction*, *3*, 43-69.

Kruijff, G. & Janicek, M. (2011) Using Doctrines for Human-Robot Collaboration to Guide Ethical Behavior, *AAAI Fall Symposium Series*, November 2011.

Lewis, P.R. & Marsh, S. (2022). What is it like to trust a rock? A functionalist perspective on trust and trustworthiness in artificial intelligence, *Cognitive Systems Research, 72,* 33-49.

Older, M. T., Waterson, P.E. and Clegg, C.W. (1997). A critical assessment of task allocation methods and their applicability, *Ergonomics*, 40, 151-172.

Roth, E.M., Sushereba, C., Militello, L.G., Diiulio, J. and Ernst, K. (2019). Function allocation considerations in the era of human autonomy teaming, *Journal of Cognitive Engineering and Decision Making*, 13, 199-220..

Stammers, R.B and Hallam, J. (1985). Task allocation and the balancing of task demands in the multi-man-machine system some case studies. *Applied Ergonomics*, 16, 251-257.

Vicente, K. J. (1999). *Cognitive work analysis: Toward safe, productive, and healthy computer-based work*, CRC press,

Waterson, P.E., Older-Gray, M. and Clegg, C.W. (2002). A sociotechnical method for designing work systems. *Human Factors*, 44, 376-391.

Yusuf, S.M and Baber, C (2022). Formalising Distributed Situation Awareness in a Team of Agents. *IEEE Transactions on Human Machine Systems, 52*, 1166-1175.